\documentclass{article}
\usepackage{latexsym}
\usepackage{graphics}
\begin{document}
\author{Mihalis Dafermos}
\title{On ``time-periodic'' black-hole solutions to certain
spherically symmetric Einstein-matter systems}
\maketitle
\bibliographystyle{plain}
\newtheorem{proposition}{Proposition}
\newtheorem{assumption}{Assiumption}
\newtheorem{theorem}{Theorem}
\begin{abstract}
This paper explores ``black hole'' solutions of various 
Einstein-wave matter systems admitting an isometry of their domain 
of outer communications taking every point to its future.
In the first two parts, it is shown that such solutions, assuming
in addition that they are spherically symmetric and the matter has a
certain structure, must be Schwarzschild or
Reissner-Nordstr\"om. Non-trivial examples of
matter for which the result applies 
are a wave map and a massive charged scalar field interacting with 
an electromagnetic field. The results thus
generalize work of Bekenstein~\cite{beken}
and Heusler~\cite{heusler} from the static to the periodic
case. In the third part, which is independent of the
first two, it is shown that 
Dirac fields preserved by an isometry of a spherically 
symmetric domain of outer communications of the type desribed above
must vanish. 
It can be applied in particular to the
Einstein-Dirac-Maxwell equations or the Einstein-Dirac-Yang/Mills
equations, generalizing
work of Finster, Smoller and Yau~\cite{3:cp}, \cite{3:ym2},
\cite{3:ym}, and also \cite{3:rn}.
\end{abstract}

For equations of evolution, time-periodic or stationary solutions
often correspond to the late time behavior of solutions for
a large class of initial data. In the general theory of 
relativity, time-periodic ``black hole'' solutions,
if they exist, seem to provide reasonable 
candidates for the final state of gravitational collapse.
Such solutions can be defined as those 
invariant with respect to an isometry of the
domain of outer communications which takes every point to
its future, or more generally, such that points sufficiently
close to infinity are mapped to their future.

In the case of a continuous family of isometries (i.e. stationary
and static solutions), this problem has a long history and goes under
the name ``no hair'' conjecture. See \cite{nohair} for a survey
of results and a recent important refinement. 
Current proofs depend on various extra assumptions
and truly satisfactory theorems have only been obtained in the
vacuum and electrovacuum static case. 

The aim of this paper is to try to generalize some results from
the static case to the spherically symmetric
``time-periodic'' case. The study of
periodic solutions to the Einstein equations was initiated in
Papapetrou \cite{papa1} \cite{papa2}; 
see also \cite{gs}. The analyses indicate that vacuum solutions 
which are periodic near null infinity should in fact be static
there, but they are far from complete, and depend very much
on analyticity assumptions on the nature of null infinity,
assumptions which do not appear to be physically valid. This paper appears
to be the first to address the issue of the existence of periodic solutions
in a non-analytic setting, in particular, in a setting compatible with 
the evolutionary hypothesis.

After briefly setting some basic assumptions (Section 1) regarding
spherical symmetry, we shall
show in Section 2 that for a 
certain class of matter, non-trivial spherically-symmetric
black-hole phenomena cannot
be described by solutions invariant with respect to a map taking
some point to 
its future. In Section 3, we shall enlarge the class
of matter for which the result applies by taking another approach,
which in effect reduces the problem to the static case.
The method of Section 3 is related to the arguments of
\cite{rigid}. 

In the spherically symmetric context, the above two sections 
generalize in particular results of 
\cite{beken} and \cite{heusler}, and Section 2, where it applies,
provides a new and easier approach for the static case. Moreover,
no assumption of invariance of the matter with respect to the isometry
is necessary, nor is any real understanding of the behavior of
the isometry on the event horizon. In fact, the results apply
equally well when the ``periodic'' assumption is weakened to
an appropriate notion of ``almost periodicity''.
Key to the results are the monotonicity properties of 
the area radius or the Hawking mass.\footnote{We prefer to use
the Hawking mass due to its special significance in spherical symmetry.
One can equally well work only with area radius.}

In Section 4, which is independent of
Sections 2 and 3, we shall show that Dirac fields preserved
by an isometry of the form described above must vanish.
The method exploits conservation
of the Dirac current. There has a been a series of recent work
\cite{3:cp}, \cite{3:ym2}, \cite{3:ym} where \emph{static}
spherically symmetric solutions of various coupled
Einstein-Dirac-matter systems are considered, and also
work \cite{3:rn} where periodic solutions of the Dirac
equation on a \emph{fixed} Reissner-Nordstr\"om background are considered.
Modulo differences in regularity assumptions, all this previous work
follows as a special case of the result of this section,
which furthermore excludes non-trivial \emph{periodic} solutions
to a large class of \emph{coupled} Einstein-Dirac-matter systems.

It should be stressed that the results of this paper
suffer from some of the deficiencies of the original ``no hair''
theorems. A critical discussion of various geometric and
regularity assumptions that appear here,
and a comparison to \cite{3:ym2}, \cite{3:ym},
\cite{3:cp} and \cite{3:rn}, is included 
in the end (Section 4).

\section{Some basic assumptions}
Let $(M,g)$ be a spacetime on which
$SO(3)$ acts by isometry and let $Q$ be the quotient manifold.
We assume that the induced metric on $Q$ has bounded curvature. 
This implies the existence of local null coordinates $u$ and $v$ 
in a neighborhood of every point of $Q$ such
that the induced metric has the form $-\Omega^2dudv$. 
Recall (\emph{e.~g.~}from~\cite{chr:sgrf}) the area 
radius~$r$ and the Hawking
mass~$m$ defined on $Q$.

Our assumption on $Q$\footnote{The extent to which the assumptions made here
can be retrieved from more primitive assumptions will be considered
in Section 4.} is basically that it strictly contain a complete
domain of outer communications ${\bf D}$, bounded in the future and past
by an event horizon, more precisely, that it contain a
region whose conformal diagram looks like:
\[
\includegraphics{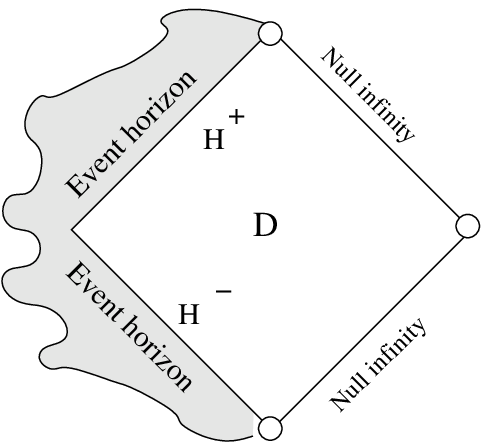}
\]
Null geodesics whose endpoint in the above diagram lies on null
infinity have in fact infinite
affine length; null infinity is thus not contained in $Q$. On
the other hand, we assume that the points on the event horizon are
contained in $Q$. The event horizon is the union of a future directed
null geodesic ray (denoted $H^+$) and a past directed ray
(denoted $H^-$). As in the diagram, we require these rays to intersect, 
\emph{i.e.}~the point $H^+\cap H^-$ is also in $Q$.\footnote{Note
that this is the familiar restrictive assumption from
the ``no-hair'' theorems; it excludes in particular
the case of the critical $e=m$ Reissner-Nordstr\"om
solution. For more, see Section 4.} Since $Q$ is open, this means that all
inextendible null rays emanating from points of ${\bf D}$
either cross the event horizon and leave ${\bf D}$ or have
infinite affine length. We also assume $r$ is bounded below
on ${\bf D}$ by a positive constant.

We will assume furthermore that the domain of outer communications
admits an isometry $\tau$, which descends to $Q$, \emph{i.~e.}, such that
it induces an isometry on ${\bf D}$ such that $\tau^*r=r$, etc. This
latter part follows for instance if we assume that the action
of $SO(3)$ on $M$ is unique.
Moreover, we will assume $\tau$ takes some point $p\in{\bf D}$ to 
its future, 
i.e.~$\tau(p)\in I^+(p)$. 

The orbit $\tau^n(p)$ must be contained in the timelike line $r=r(p)$.
Moreover, since the distance between $\tau(p)$ and $p$ is non-zero and
must equal to the distance between $\tau^{(n+1)}(p)$ and $\tau^n(p)$,
it follows that $\tau^n(p)$ approaches future timelike infinity,
and $\tau^{-n}(p)$ approaches future past timelike infinity. In particular,
given any point $q\in{\bf D}$, there will be a $\tau^i(p)$ in $I^+(q)$,
and a $\tau^{-i}(p)$ in $I^-(q)$. Since the causal relation between two
points is preserved by an isometry, it follows that $\tau(q)$ and
$q$ cannot be connected by an achronal curve. For $\tau^{2i}q$ could
then never be in the future of $\tau^i(p)$. Thus $\tau(q)\in I^+(q)$
for all $q\in {\bf D}$, and moreover, there are no limit points of
the orbits $\tau^n(q)$ in the closure of ${\bf D}$.

The key behind all our arguments will be to show that the
existence of the isometry implies that certain quantities vanish
on the event horizon. In the
simplest cases, covered by the
following section, this will then uniquely determine the
wave matter to be constant on the event horizon, and then constant
everwhere, by
application of a uniqueness theorem
for solutions of the characteristic
initial value problem.\footnote{Note that in the spherically symmetric
context, one can interchange the notions of spacelike and
timelike on the quotient manifold $Q$, making ${\bf D}=J^+(H^+\cup H^-)$,
and thus, for
appropriate equations, data on the
event horizon determines solutions throughout ${\bf D}$, by applying
standard theorems~\cite{char}.} 
In the case
where the matter satisfies the weak energy condition, the vanishing
of this quantity is proven using the monotonicity
properties of the Hawking mass.

\section{Exploiting the coupling with gravity}

We refer the reader to~\cite{chr:sgrf} for a derivation of the Einstein
equations in spherical symmetry with a general energy-momentum tensor.
We will assume here that these equations are satisfied pointwise
(\emph{i.~e.}~all functions that appear are bounded)
in the null coordinate charts of our atlas for an induced energy momentum 
tensor $T_{ab}$ on $Q$ which satisfies the energy condition
$T_{uu}\ge0$, $T_{uv}\ge0$, $T_{vv}\ge0$. Here we always select
$v$ such 
that null geodesic rays from points of ${\bf D}$
generated by $\partial_v$ are future-directed
and have infinite affine length (i.~e.~``terminate'' on
null infinity). It follows
from
\begin{equation}
\label{requ}
\nabla_a\nabla_br=\frac{1}{2r}(1-\partial^cr\partial_cr)g_{ab}
-r(T_{ab}-g_{ab}{\rm tr}T)
\end{equation}
that $\partial_u r\le0$ and 
$\partial_v r\ge0$ in ${\bf D}$\footnote{The arguments
are similar to those of~\cite{chr:sgrf}. 
Assuming one of the inequalities does not hold,
one argues by integrating
$(\ref{requ})$ that $r$ will have to become zero after a finite
affine length in the direction of a null geodesic that
terminates at the event horizon, a contradiction.}
and then, from
\begin{equation}
\label{massequation}
\partial_am=r^2(T_{ab}-g_{ab}{\rm tr}T)\partial_br
\end{equation}
that $\partial_u m\le0$ and $\partial_v m\ge0$.

We then have the following
\begin{proposition}
Let $\tau$ be an isometry
of the domain of outer communications ${\bf D}$ as described
in the previous section.
It follows that $\partial_vr=0$ and 
$\partial_vm=0$ on
$H^+$ and $\partial_ur=0$, $\partial_um=0$ on $H^-$. 
\end{proposition}

\noindent{\it Proof.} 
The proof is by contradiction. Suppose $p$ and $q$ are two points
on $H^+$ such that $r(q)=r(p)+\epsilon$ for $\epsilon>0$. By continuity
of $r$ there exists
a point $q'\in {\bf D}$ on the ray generated by $-{(\partial_u)}_q$ 
such that $r(q')>
r(p)$. 
\[
\includegraphics{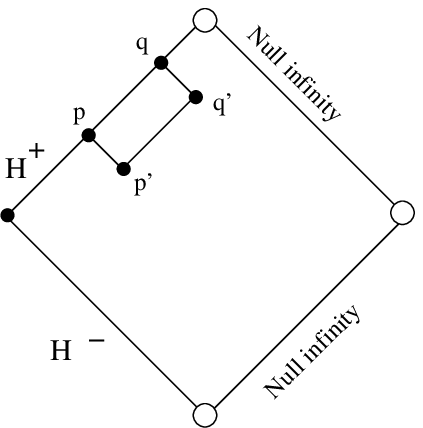}
\]
It follows from the equations $\partial_ur\le0$, $\partial_vr\ge0$ that
$r>r(p)$ in ${\bf D}\cap J^+(q')$. Now consider the point
$p'$ at the intersection of the null
ray generated by 
$-{(\partial_v)}_{q'}$ and the null ray generated by $-{(\partial_v)}_p$.
Again, by the relation $\partial_ur\le0$ it follows that
$r(p')\le r(p)$. Now the assumption on $\tau$ implies that there 
exists an $N$ such that
$\tau^n(p')\in{\bf D}\cap J^+(q')$ for all $n>N$. But since
$\tau$ is an isometry $r(\tau^n(p'))=r(p')\le r(p)$. This is a
contradiction. One can then apply the same argument with $m$
in place of $r$, and then for $H^-$ replacing $H^+$,
thus completing the proof. $\Box$

In virtue of the equation $(\ref{massequation})$ and the boundedness
of $T_{uv}$, it follows from the above
proposition that since $\partial_vm=0$ and $\partial_vr=0$ on $H^+$ and
similarly $\partial_u m=\partial_ur=0$ on $H^-$, we have $T_{vv}=0$ 
on $H^+$ and $T_{uu}=0$ on $H^-$.

We now proceed to outline the more restrictive
assumptions on the structure of the matter which will
be necessary for our results. The first set
of assumptions reflects the structure of the
energy-momentum tensor itself.  These are:
\begin{enumerate}
\item
$T=T(\Psi, F, g)$ where $F$ is a skew symmetric $2$-tensor,
and $\Psi$ takes values in some space endowed with a connection 
$\tilde\nabla$, 
such that if $\tilde\nabla_X\Psi=0$ identically for
all $X\in T^*M$,
then $T$ corresponds to the
energy momentum tensor of a spherically symmetric electric field
$F_{\mu\nu}$
satisfying the source-free Maxwell equations. Here 
$\tilde\nabla_X$ is the induced connection on $M$.
\item
$T_{vv}=0$ should imply $\tilde\nabla_v\Psi=0$,
and $T_{uu}=0$ should imply $\tilde\nabla_u\Psi=0$.
Here $\tilde\nabla_v=\tilde\nabla_{\partial_v}$, etc.
\end{enumerate}
One example of such a $T$ is the energy momentum tensor
generated by a wave map $\phi:(M,g)\to (N,h)$ interacting 
(via the gravitational field only, since it does not carry charge) in
an electromagnetic field $F_{\mu\nu}$:
\[
T_{\mu\nu}=
F_{\mu\lambda}F_{\nu\rho}g^{\lambda\rho}
-\frac1{4}g_{\mu\nu}F_{\lambda\rho}F_{\sigma\tau}
g^{\lambda\sigma}g^{\rho\tau}
+h_{AB}(\phi^A_{;\mu}\phi^B_{;\nu}-
\frac12 g_{\mu\nu}g^{\rho\sigma}\phi^A_{;\rho}\phi^B_{;\sigma})
\]
A special case of the above is of course when $N$ is $R^n$
with the flat metric and $\phi$ is then a collection of $n$
real scalar fields.
Another example is the energy momentum tensor generated
by a massive complex scalar field $\phi$ in an electromagnetic
field $F_{\mu\nu}$ with electromagnetic potential $A_\mu$ 
\begin{eqnarray*}
T_{\mu\nu}&=&F_{\mu\lambda}F_{\nu\rho}g^{\lambda\rho}
-\frac14g_{\mu\nu}F_{\rho\sigma}F_{\lambda\kappa}
g^{\rho\lambda}g^{\sigma\kappa}\\
&&-\frac12 g_{\mu\nu}M^2\phi\bar{\phi}+\frac12(\phi_{;\mu}\bar{\phi}_{;\nu}+
\bar{\phi}_{;\mu}\phi_{;\nu})\\
&&+\frac12(-\phi_{;\mu}ieA_\nu\bar{\phi}+
\bar{\phi}_{;\nu}ieA_\mu\phi+\bar{\phi}_{;\mu}ieA_\nu\phi-
\phi_{;\nu}ieA_{\mu}\bar{\phi})\\
&&-\frac12g_{\mu\nu}g^{\rho\sigma}
(\phi_{;\rho}+ieA_\rho\phi)(\bar{\phi}_{;\sigma}-
ieA_{\sigma}\bar{\phi})+e^2A_{a}A_{b}\phi\bar{\phi}.
\end{eqnarray*} 

For $T$ satisfying $1$ and $2$, it follows that
$\tilde\nabla_v\Psi=0$ on $H^+$ and $\tilde\nabla_u\Psi=0$ on $H^-$. Since
the restriction of a connection to a one-dimensional set
is trivial, we can then choose coordinates for a space representing
the degrees of freedom for
$\Psi$ such that $\Psi$ is in fact constant on the 
event horizon. If in local
coordinates $x_a\in Q$ 
the system of equations for $\Psi$, with $F_{\mu\nu}$,
$g_{ab}$, and $r$ fixed, is of the form
\begin{equation}
\label{waveE}
\partial^a\partial_a\Psi=F(\nabla\Psi,\Psi, x_a)
\end{equation}
with $F$ a sufficiently regular function\footnote{Note that for fields
which couple directly to the $F_{\mu\nu}$ tensor, there is a 
regularity assumption on $F$ as well as on $g$ implicit
in $(\ref{waveE})$.},
then the characteristic initial value problem with initial data on the
event horizon is locally well posed, provided $\Psi$ is assumed
sufficiently regular~\cite{char}.\footnote{Recall the comment
from before that to apply $\cite{char}$,
one should first redefine the metric $g_{ab}$ to be its negative,
so ${\bf D}$ becomes $J^+(H^+\cup H^-)$.} 
If this equation admits the solution
$\Psi=\Psi(H^+\cup H^-)$, then this must be the only solution 
in the vicinity of the horizon, and by
a continuity argument, this domain of dependence property
can be extended to guarantee uniqueness throughout ${\bf D}$.
A sufficient condition for this is clearly $F(0,\Psi,x_a)=0$.
Thus, in view of the fact
that spherically symmetric solutions of the Einstein-Maxwell
equations are necessarily Reissner-Nordstr\"om, we have
\begin{theorem}
If $T_{\mu\nu}$ satisfies conditions 1 and 2, and
$\Psi$ satisfies a system of the form
$(\ref{waveE})$, with $F(0,\Psi,x_\alpha)(p)=0$,
and if $\tau$ is as in Proposition 1,
it follows that ${\bf D}$ coincides with
the domain of outer communications of a Reissner-Nordstr\"om
solution.
\end{theorem}

Note that the above theorem applies to the wave map system,
which can be written
\[
\partial^{\alpha}\partial_\alpha\Phi=\Gamma(\Phi)(|\nabla\Phi|^2).
\]
where $\Gamma$ is an expression involving the Christoffel symbols
of $(N, g)$. (Compare with \cite{heusler}. The above argument
reproves, in particular, the static result, and seems considerably
easier, as it does not depend on the geometry of the target.)
Also note that, in the above argument,
we have not assumed that $\tau$
preserves $\Psi$, only that it preserves the metric. In fact,
it suffice to assume that given any point $p$, then for
all $\epsilon$ there exists an $N(\epsilon, p)$ such that
$|m(\tau^n(p))-m(p)|<\epsilon$ for $|n|\ge N$. Such solutions
can be called ``almost periodic''.

\section{Constructing a Killing vector and reducing to the static case}

Unfortunately, as it stands, the argument of the previous
section cannot be applied in the
case of a complex scalar field or a massive scalar field, for
then the dependence of $F$ on $\Psi$ is not of the type described above.
In particular, there do not exist constant non-zero solutions.

It is perhaps instructive to compare here with the
static case.
The argument of Bekenstein~\cite{beken}, 
say for the scalar field $\Box\phi=M\phi$ with
$M>0$, goes roughly as follows: Integrating 
the equality $\nabla_\alpha(\phi\nabla^\alpha\phi)=
\nabla_a\phi\nabla^\alpha\phi+M\phi^2$ using Gauss's
theorem, the boundary contributions along the
event horizon vanish, while the contributions along two spacelike
curves which are carried to one another by the isometry cancel.
Moreover, the divergence is non-negative, since in a static
solution $\nabla\phi$ is spacelike, and $0$ only if the solution vanishes.
Thus, either the solution is identically $0$, 
or there must be a boundary contribution
at infinity, i.e.~the solution does not decay
as $r\to\infty$. (This would imply that the curvature does not decay,
and thus the solution would not be asymptotically flat.)

The above Bochner-type method and arguments based on it
cannot be applied directly in the 
periodic case as $\nabla\phi$ may
have negative length. With slightly more effort than in 
Section 2, one can show that for various examples of matter--including
the case of a charged massive
scalar field, for instance--our 
initial data determine a static solution, and then
apply the above argument to show that this solution
must thus not decay at infinity.

The idea is similar in spirit to Theorem 1, except that
now we will apply the uniqueness theorem to the solution
of the characteristic value problem to a system
of second order hyperbolic
equations derived from Killing's equation. 

First, we introduce the following new assumptions:
Letting
$x^A$, $x^B$ denote coordinates on $S^2$,
we assume

\begin{enumerate}
\item
$\partial_vr=0$ on $H^+$ and $\partial_ur=0$ on $H^-$
\item
$T_{vv}=0$, $\nabla_uT_{vv}=0$ on $H^+$ and 
$T_{uu}=0$, $\nabla_vT_{uu}=0$ on $H^-$.
\item
$\nabla_vT_{AB}=0$ on $H^+$ and $\nabla_uT_{AB}=0$ on $H^-$.
\end{enumerate}

Given an isometry $\tau$ as before,
Assumptions 1, 2 and 3 above follow for a large class of matter,
including the case of a 
complex scalar field interacting in an electromagnetic field.
(See the Appendix.)

We define $v$ and $u$ on the event horizon so as
to yield an affine distance
on the event horizon as measured from the point $H^+\cap H^-$
on $H^+$ and $H^-$ respectively, i.e.~we will be assuming
that $g_{uv}=-1$ on $H^+\cup H^-$. 
We now will define a particular
null vector field $K$ on the event horizon,
and extend it
to ${\bf D}$ as the unique solution
of the initial value problem, with initial data on the event
horizon\footnote{Recall the comment in Section 1 about
the well-posedness of this problem in spherical symmetry, because
of the symmetry between timelike and spacelike directions.}, 
for the equation
\begin{equation}
\label{kileq}
\Box K^{\alpha}=-K^\beta R_{\beta\gamma}g^{\gamma\alpha}.
\end{equation}
The choice of the definition will be to ensure that
$L_K g_{\mu\nu}=0$ on $H^+\cup H^-$. For now write
$K|_{H^+}=K^v(v)\partial_v$ and $K|_{H^-}=K^u(u)\partial_u$,
where we will determine immediately following what
$K^v(v)$ and $K^u(u)$ have to be.

Let us concentrate first on $H^+$. In the null coordinates
defined above (where in addition $x^A$ and $x^B$ are taken to be normal
coordinates), the only non-vanishing Christoffel symbol
are $\Gamma^u_{uu}$, $\Gamma^B_{uA}$ and $\Gamma^v_{AB}$. 
Outside of $H^+$, $\Gamma^u_{uu}$,
$\Gamma^v_{vv}$, $\Gamma^B_{uA}$, $\Gamma^B_{vA}$, $\Gamma^u_{AB}$, 
$\Gamma^v_{AB}$ are the only 
non-vanishing components.
Note also that on $H^+$,
\begin{eqnarray*}
R&=&2g^{uv}R_{uv}+g^{AB}R_{AB}\\
 &=&-2(-\partial_u\Gamma^v_{vv}-\partial_u\Gamma_{vA}^A)
+g^{AB}(\partial_u\Gamma^u_{AB}+\partial_v\Gamma^v_{AB})\\
&=&2\partial_u\Gamma^v_{vv}+4g^{AB}\partial_u\partial_v g_{AB}\\
&=&2\partial_u\Gamma^v_{vv}+2(R+2R_{uv}),
\end{eqnarray*}
and thus, we have that
\[
\partial_u\Gamma_{vv}^v=-\frac12R-2R_{uv}.
\]
We compute 
\[
\Box K^u=-2\partial_u\partial_vK^{u},
\]
\[
\Box K^v=-2\partial_u\partial_vK^v
	 +\partial_vK^v(-g^{AB}\Gamma_{AB}^v)+K^v(-\partial_u\Gamma_{vv}^v),
\]
and thus
$(\ref{kileq})$ gives,
\begin{equation}
\label{toeva}
\partial_u\partial_vK^u=0,
\end{equation}
\begin{eqnarray}
\label{toallo}
\nonumber
\partial_v\partial_uK^v&=&-\frac{1}2\partial_vK^vg^{AB}\Gamma_{AB}^v
-\frac12(\partial_u\Gamma_{vv}^v+R_{uv})K^v\\
\nonumber
&=&-\frac{1}2\partial_vK^vg^{AB}\Gamma_{AB}^v
+\frac14(R+2R_{uv})K^v\\
&=&-\frac14\partial_vK^v\int_0^v{(R+2R_{uv})dv}+\frac14K^v 
(R+2R_{uv}).
\end{eqnarray}

Recalling $(L_Kg)_{\alpha\beta}=K_{\alpha;\beta}+K_{\beta;\alpha}$,
in view of our knowledge of the Christoffel symbols, and
the fact that $K_v=0$ on $H^+$,
we obtain
\begin{equation}
\label{enapom1}
(L_Kg)_{uv}=\partial_uK_v+\partial_vK_u=
-\partial_uK^u-\partial_vK^v,
\end{equation}
\[
(L_Kg)_{vv}=2\partial_vK_v=-2\partial_uK^u=0,
\]
\[
(L_Kg)_{vA}=0, (L_Kg)_{uA}=0,
\]
\[
(L_Kg)_{AB}=-\Gamma^{u}_{AB}K_u-\Gamma^v_{AB}K_v=0,
\]
\begin{equation}
\label{enapom2}
(L_Kg)_{uu}=2(\partial_uK_u-K_u\Gamma_{uu}^u)=-2\partial_uK^v.
\end{equation}
Thus, if we are to have $(L_Kg)_{\alpha\beta}=0$ on $H^+$, it follows
from $(\ref{enapom1})$ that
\begin{equation}
\label{haber}
\partial_uK^u=-\partial_vK^v.
\end{equation}
Rewriting $(\ref{toeva})$ as $\partial_v\partial_uK^u=0$,
it follows from $(\ref{haber})$ that
\[
\partial_v\partial_vK^v=0, 
\]
and thus that $K^v=Cv$. Using $(\ref{haber})$ again, and the
same argument, it follows that $K^u=-Cu$ on $H^-$.

Of course, to show that indeed we have $(L_Kg)_{\alpha\beta}=0$
on $H^+$, for the $K$ defined above, it remains to show, in view
of $(\ref{enapom2})$, that
\[
\partial_uK^v=0.
\]
Since the above equation is indeed true at $H^+\cap H^-$,
it follows that it is enough to show that $\partial_v\partial_uK^v=0$,
or, by $(\ref{toallo})$, that
\[
-\frac14\partial_vK^v\int_0^v{(R+2R_{uv})dv}+\frac14K^v 
(R+2R_{uv})=0.
\]
Assumption 2 together
with the conservation of energy-momentum implies that
$\partial_vR_{uv}=0$, and Assumption 3 implies that
$\partial_vR=0$, and thus, $R+2R_{uv}=c$ where
$c=(R+2R_{uv})|_{H^+\cap H^-}$.
Thus we compute
\[
-\frac14\partial_vK^v\int_0^v{(R+2R_{uv})dv}+\frac14K^v 
(R+2R_{uv})
=-\frac14C(cv)+\frac14(Cv)c=0.
\]
We can take then $C=1$ and we have found a nontrivial
vector field vanishing at $H^+\cap H^-$ satisfying
$L_Kg=0$ on $H^+$, and similarly,
$L_Kg=0$ on $H^-$ as well.

Denote now the totality of matter by $\Phi$.
In terms of this $K$ defined, we assume further
\begin{enumerate}
\item[4]
$L_K\Phi=0$\footnote{The expression $L_K\Phi$ can be
tricky to define if the equations have a gauge invariance. Typically,
this will mean that there is a choice of gauge for which the
matter can be expressed by some $\Phi$ for which $L_K\Phi=0$. 
See the appendix for the case of a complex scalar field.}
 on $H^+\cup H^-$.
\item[5]
The quanitities
$L_K g_{\mu\nu}$ and
$L_K\Phi$ satisfy a system of equations which,
when everything else is treated as fixed,
only admits the zero solution if they vanish
on $H^+\cup H^-$.
\end{enumerate}

All our assumptions taken together imply that we have produced
a vector field $K$ such that
$L_Kg=0$, i.e.~a Killing field $K$ on ${\bf D}$. 
Note that a similar argument ensures
that $K$ is also a Killing field ``downstairs'', i.e.~that $Kr=0$.
From this, it follows that $K$ must
be timelike. Since $K$ does not vanish identically on the
event horizon, it follows that there exists a $p\in{\bf D}$ such 
that $K(p)\ne0$. Thus $K$ does not vanish along the line
$r=r(p)$, which must be the orbit $\phi_t(p)$ where
$\phi_t$ denotes the one parameter group of isometries
generated by $K$. Since all future directed constant-$u$ null rays
must intersect the line $r=r(p)$, it follows that $K$ can nowhere
vanish. For if it did at some point $q$, then choosing a point $s$
on $r=r(p)$ which can be connected to $q$ by a spacelike curve,
then $\phi_t(s)$ for large enough $t$ is in the future of $\phi_t(q)=q$,
which contradicts the fact that $\phi_t$ is an isometry.

We have thus proven
\begin{theorem}
For an Einstein-matter system satisfying Assumptions 1, 2, 3, 4,
and 5
above, the domain of outer communications
is static.
\end{theorem}

\section{Exploiting a conserved current: the case of the Dirac equation}
In the case of Dirac fields, the arguments outlined 
above do not apply because this matter
does not satisfy the positive energy condition. This is related to the
fact that the Dirac field probably provides a reasonable model only after
second quantization. But in fact,
considerations regarding periodic solutions are even easier than
in the previous section, 
and can be studied without applying the coupling
with gravity, which played a central role in the previous argument.

We refer the reader to \cite{3:rn} for background on this
problem in the uncoupled case, and to \cite{3:cp}, \cite{3:ym2},
\cite{3:ym} in the coupled static case. In particular, we recall
the Dirac matrices $G^\alpha$, which operate on Dirac
fields $\Psi$, which are sections of an appropriate spinor
bundle. The precise
form of the Dirac equation will depend on the other matter
fields present to which the field is coupled.
We will simply assume that
$\Psi$ satisfies in local coordinates,
after fixing the metric and the other matter fields, a
\emph{linear} equation of the form
\begin{equation}
\label{dirac}
iG^\alpha\partial_\alpha\Psi=F(\Psi),
\end{equation}
where $F(0)=0$. Note that by squaring the Dirac operator,
it follows that $\Psi$ satisfies a system
\begin{equation}
\label{dirac2}
\Box\Psi=\tilde{F}(\nabla\Psi,\Psi,x_\alpha).
\end{equation}
We further assume that the
vector field $\bar\Psi G^\alpha\Psi$ provides a positive
current, i.e.~
\[
\bar\Psi G^\alpha\Psi X^\beta g_{\alpha\beta}\ge 0
\]
when $X^\beta$
is future directed and timelike, with equality only in the case where 
$\Psi$ vanishes, and moreover, this current is conserved:
\begin{equation}
\label{conserved}
\nabla_\alpha(\bar\Psi G^\alpha\Psi)=0.
\end{equation}

We do not assume that $\Psi$ is spherically symmetric, but we do assume
that it is defined on a spherically symmetric domain of outer communications as before, preserved by an isometry
$\tau$, as before, in the sense that
\begin{equation}
\label{per=}
\tau_*(\bar{\Psi}G^\alpha\Psi)=\bar\Psi G^\alpha\Psi.
\end{equation}

Moreover, we assume that all other matter fields are spherically
symmetric, and thus we can write $\Psi$ as the sum of
spherical harmonics each of which satisfy a wave equation of
the form $(\ref{dirac2})$, but on $Q$, not on $M$.

Consider a spacelike curve $\gamma$ which divides ${\bf D}$ into two connected
components, and intersects the event horizon at $H^+\cap H^-$.
Let $X$ be the future normal vector field to $\gamma$.
Fix a point $q$ on the event horizon and $p$ on $\gamma$. We denote
the part of $\gamma$ connecting $p$ with spacelike infinity
by $\gamma_p$. Then for
an isometry $\tau$ as in Proposition 1, There exists an $n$
such that $\tau^n(p)$ and $q$ can be connected by a spacelike
curve $\tilde{\gamma}$:
\[
\includegraphics{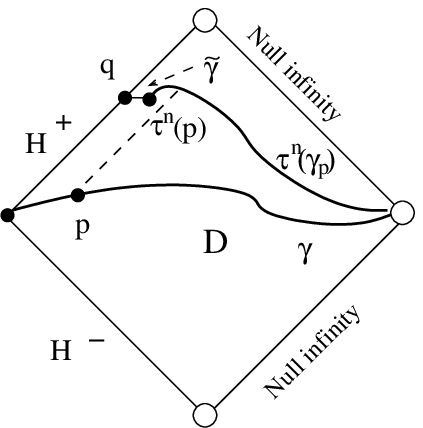}
\]
We will assume $\Psi$ is locally bounded, and
that 
\[
\int_{\gamma}{\bar\Psi G^\alpha\Psi X^\beta g_{\alpha\beta}}<\infty.
\]
The latter assumption is quite reasonable in view of the
fact that this integral should equal to the probability
of observing the particle on $\gamma$, which should be normalizable
to something less than $1$.
Now integrating the conservation law $(\ref{conserved})$
and applying Gauss' theorem,
and since
\[
\int_{\tilde{\gamma}}{\bar\Psi G^\alpha\Psi X^\beta g_{\alpha\beta}}>0
\]
it follows that 
\begin{eqnarray*}
\int_{H^+\cap J^-(q)}{\bar\Psi G^\alpha\Psi X^\beta g_{\alpha\beta}}
&\le&
\int_\gamma{\bar\Psi G^\alpha\Psi X^\beta g_{\alpha\beta}}
-\int_{\tau^n{(\gamma_p)}}
{\bar\Psi G^\alpha\Psi (\tau^n_*X)^\beta g_{\alpha\beta}}\\
&=&
\int_\gamma{\bar\Psi G^\alpha\Psi X^\beta g_{\alpha\beta}}
-
\int_{\gamma_p}{\bar\Psi G^\alpha\Psi X^\beta g_{\alpha\beta}}\\
&=&\int_{\gamma\backslash\gamma_p}
{\bar\Psi G^\alpha\Psi X^\beta g_{\alpha\beta}},
\end{eqnarray*}
where the second line follows from the fact
that $\tau$ is an isometry and $(\ref{per=})$.

But as $p\to H^+\cap H^-$, the term on the right hand
side approaches $0$.
Thus, since the left hand side is nonnegative,
it follows that
\[
\int_{H^+\cap J^-(q)}{\bar\Psi G^\alpha\Psi }N^\beta g_{\alpha\beta}=0
\]
for all $q$ and consequently,
$\bar\Psi G^\alpha\Psi N^\beta g_{\alpha\beta}=0$ 
identically on $H^+$, and similarly,
$\bar\Psi G^\alpha\Psi N_-^\beta g_{\alpha\beta}=0$ on
$H^-$, where $N_-$ denotes the null vector tangent to 
$H^-$. Since $N_-+N$ at $H^-\cap H^+$ is timelike,
it follows by the positivity of the current that
$\Psi$ in fact vanishes there.

It turns out that the behavior of $\Psi$ on the event
horizon, deduced above, together with the Dirac equation,
imply that $\Psi$ vanishes identically on the event horizon:

Choose coordinates $u$, $v$, $x^1$,
and $x^2$ in a neighborhood of $H^+\cap H^-$, such that,
$g=-\Omega^2dudv+\tilde{g}_{ij}dx^idx^j$. It follows
from the properties deduced above that a spinor representation
can be chosen such that $G^u\Psi=0$, $G^u\partial_v\Psi=0$,
and $G^u\partial_{x^i}\Psi=0$ on $H^+$, while
$G^v\Psi=0$, $G^v\partial_u\Psi=0$,
and $G^v\partial_{x^i}\Psi=0$ on $H^-$.

From the anticommutation relations it follows
that $G^vG^v=0$, $G^uG^u=0$.
Multiplying the Dirac equation $(\ref{dirac})$ by $G_u$,
and restricting to $H^+$,
one obtains,
\[i(G^uG^v\partial_v\Psi+G^uG^{x^i}\partial_{x^i}\Psi)=G^u(F(\Psi)).\]
Since $G^uG^{x^i}=-G^{x^i}G^u$ by the anticommutation relations,
it follows from $G^u\Psi=0$ that
$iG^uG^v\partial_v\Psi=G^uF(\Psi)$.
Again, from the anticommutation relations, one obtains that
$G^uG^v=2g^{uv}-G^vG^u$, and thus, since $G^u\partial_v\Psi=0$,
\begin{equation}
\label{ode}
\partial_v\Psi=\tilde{f}(\Psi)
\end{equation}
for a well-behaved function
$\tilde{f}$ with $\tilde{f}(0)=0$.

From the fact shown above that $\Psi=0$
at $H^+\cap H^-$, it now follows immediately 
from $(\ref{ode})$ that $\Psi$ must
vanish identically on $H^+$. One argues in the same way to
obtain that $\Psi$ vanishes identically on $H^-$.

This condition completely determines initial data for
the characteristic initial value problem for each of
the spherical harmonics of the Dirac 
equation\footnote{See the remark in the previous section about
the change of sign of the metric $g_{ab}$}, 
and thus assuming that $\Psi$ is in a space 
sufficiently regular, all the spherical harmonics
must vanish identically in ${\bf D}$ by uniqueness
of the solution of the characteristic initial value problem,
and thus $\Psi=0$ in ${\bf D}$.

Again, it is clear from the proof that one can replace the assumptions
on $\tau$ with the assumption that for all $p$, $\epsilon$, there
exist $N(\epsilon,p)$ such that $|n|\ge N$ implies
\[
\left|
\int_{\gamma_p}{\bar\Psi G^\alpha\Psi X^\beta g_{\alpha\beta}}
-
\int_{\gamma_{\tau^n(p)}}{\bar\Psi G^\alpha\Psi X^\beta g_{\alpha\beta}}
\right|<\epsilon.
\]
There is thus a sense in which the result holds for ``almost
periodic solutions'' as well.

\section{A note on the assumptions}

As discussed in the beginning, the motivation for
considering ``time-periodic'' solutions $(Q,g)$ is as ``limiting''
final states of graviational collapse. Thus, \emph{a priori},
it makes sense only to assume that $Q$ be defined to the
future of a spacelike surface $S$.
\[
\includegraphics{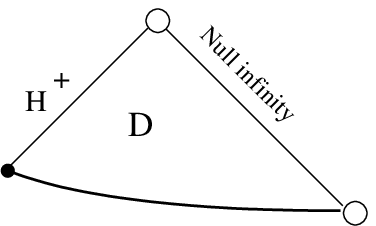}
\]

In view of our assumptions on the existence of an isometry
$\tau$, however, given a fundamental domain $F$ such that 
$S\subset\partial{F}$, one can construct in an obvious 
way $\tilde{Q}$ and $\tilde{\tau}$ an isometry of 
$\tilde{Q}$, such that $\tilde{Q}=Q\cup_i\tau^{-i}F$.
In the spherically symmetric case, if the energy momentum
tensor satisfies the energy condition, it
follows that this spacetime will also have a past boundary.
For it is clear by the arguments of Section 2 that since
$\partial_vr>0$ in ${\bf D}$, it must become less than
the infimum of $r$ within finite affine length in the
direction $-\partial_v$. Moreover, by arguments similar in spirit
to the proof of Proposition $2$, this boundary can be shown
to have a natural null structure and we can denote it as before
by $H^-$.
\[
\includegraphics{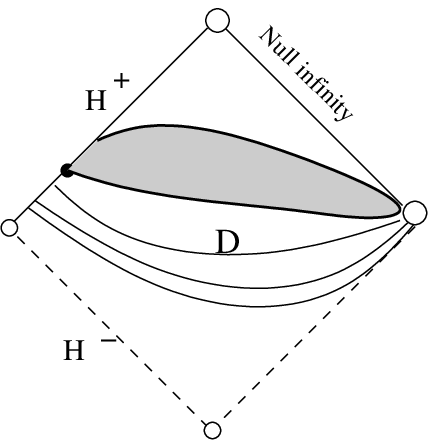}
\]

On the original $Q$, it is reasonable to assume only the regularity
that might be induced from being a limit of regular ``collapsing''
spacetimes outside the event horizon $H^+$. By monotonicity,
$r$ and $m$ can be extended to $H^+$, but the regularity
of matter fields may be quite weak. The regularity implications
for $H^-$ may in fact be weaker.

Moreover, there is no guarantee that the spacetime can be
extended so that $H^+$ and $H^-$ intersect, much less for fields
to have any sort of regularity there. (Compare with the 
Reissner-Nordstr\"om solution with $e=m$.)

Our results, on the other hand, depend very much on
the existence of the point $H^+\cap H^-$, and on some 
assumptions of regularity along these null curves. While both
of these aspects of the setup could possibly be weakened,
it is clear that some sort of regularity assumption will
certainly be an integral part of any argument involving
well-posedness of the characteristic value problem.

For another approach to these issues of regularity, it
is instructive to compare with the work of Finster, Smoller,
and Yau.
In \cite{3:ym2}, \cite{3:ym}, and \cite{3:cp}, static 
spherically symmetric solutions to
the Einstein-Dirac-Yang/Mills and
Einstein-Maxwell-Dirac equations were considered, and in 
\cite{3:rn}, periodic solutions of the Dirac equation on
a Reissner-Nordstr\"om background. 

In the above series of papers, solutions of the
Dirac equation are in fact permitted to blow up along the event horizon,
in a very specific way, and the normalization condition is relaxed
near the event horizon.\footnote{The motivation for
this seems to be more the global behavior of the special $(r,t)$ coordinate system and its associated gauge, than observations
which could be made by local observers, employed here.} Thus, from one
point of view, their assumptions could be considered weaker.
On the other hand, \cite{3:ym2}, \cite{3:ym} and \cite{3:cp} 
introduce assumptions at the level of $C^\infty$ of the metric 
at the horizon, and various auxiliary coordinated-dependent
conditions and power-law assumptions,
while \cite{3:rn} depends very much on
an assumption on the vanishing 
of a certain flux over $H^-$. In \cite{3:rn},
the considerations regarding the
analysis of $\Phi$ on $H^+$ relate to a particular choice 
of ``extension'' of $\Phi$ beyond the event horizon, and they 
also seem to depend on the conformal geometry of the interior 
of the Reissner-Nordstr\"om black hole, a geometry which 
is thought to be unstable. It is unclear how any of these conditions 
are justified if we view $Q$ as ``generated'' by the process
outlined in the beginning of this section, and thus whether
anything is gained by allowing \emph{a priori} for a more singular
behavior of $\Phi$.

\section{Acknowledgment}

I thank Piotr Chru\'sciel for some very useful discussions on a preliminary
version of this paper.

\section{Appendix}

We will show that a complex scalar field indeed satisfies
the assumptions of Section 3. To reduce the equations to a determined
system, we will have to set a gauge. We will require thus
that $A_v=0$ on $H^+$ and $A_u=0$ on $H^-$, and the components
$A_B=0$ as well, where $x^B$ are coordinates on $S^2$.
We will also introduce the notation $D$ for the
covariant derivative defined by the connection $A$, i.e.~we
have $D_{\mu}\phi=\phi_{,\mu}+ieA_{\mu}\phi$.

Note that the only non-vanishing components of the electromagnetic
tensor $F_{\mu\nu}$ are $F_{uv}$ and the collection $F_{AB}$, where
$A$ and $B$ range over coordinates on $S^2$. 

Thus,
\begin{equation}
\label{comp1}
T_{vv}=D_v\phi \overline{D_v{\phi}},
\end{equation}
\begin{equation}
\label{comp2}
T_{uu}=D_u\phi \overline{D_u{\phi}},
\end{equation}
\[
T_{uv}=-\frac14g_{uv}F_{AB}F_{CD}g^{AC}g^{BD}
-\frac12g_{uv}(D_u\phi\overline{D_v{\phi}}+
D_v\phi\overline{D_u{\phi}}),
\]
\[
T_{AB}=F_{AC}F_{BD}g^{CD}-\frac14g_{AB}F_{MN}F_{CD}g^{MC}g^{ND}.
\]

In particular, the existence of $\tau$ implies that $T_{vv}=0$ on 
$H^+$, $T_{uu}=0$ on $H^-$, and thus $\partial_v\phi=D_v\phi=0$
on $H^+$ and $\partial_u\phi=D_u\phi=0$
on $H^-$. Moreover, since $\nabla_uT_{vv}=\partial_uT_{vv}$,
applying $\partial_u$ to $(\ref{comp1})$ yields
$\nabla_uT_{vv}=0$ on $H^+$ and similarly $\nabla_vT_{uu}=0$
on $H^-$. Assumptions 1 and 2 of Section 3 thus hold.

Now, Maxwell's equations
\[
F_{\mu\nu;\rho}g^{\nu\rho}-ie\phi\overline{D_\mu{\phi}}+
ie\overline{\phi}D_\mu\phi=0,
\]
restricted to $H^+$, yields the equation
\[
\partial_vF_{vu}=0,
\]
and on $H^-$, the equation 
\[
\partial_uF_{vu}=0.
\]
Similarly, the equation
$F_{[AB,v]}=0$ yields
\[
\partial_vF_{AB}=0,
\]
and
\[
\partial_uF_{AB}=0,
\]
throughout ${\bf D}$. 
Thus we have $\partial_vT_{AB}=0$ on $H^+$ and
$\partial_uT_{AB}=0$ on $H^-$, and this implies Assumption 3.

To write a determined system of equations, we impose
the equation
$\nabla^\alpha A_\alpha=0$.
Note that this equation, together with the condition that
$A_u=0$ on $H^+$ and $A_v=0$ on $H^-$ implies that 
$L_KA_\mu=0$ on $H^+\cup H^-$.
For, on $H^+$ we obtain that $A_{u,v}=\frac12 F_{uv}$,
and thus
\begin{eqnarray*}
(L_KA)_u&=&K^\mu A_{\mu,u}+K^{\mu}_{,u}A_\mu\\
&=&K^vA_{u,v}+K^u_{,u}A_u\\
&=&K^vA_{u,v}-K^v_{,v}A_u\\
&=&\frac12K^vF_{uv}-\frac12\partial_vK^vF_{uv}v\\
&=&\frac12CvF_{uv}-\frac12CvF_{uv}=0,
\end{eqnarray*}
\[
(L_KA)_v=K^vA_{v,v}+K^u_{,v}A_u+K^v_{,v}A_v=0+0+0=0.
\]

Our equations for the matter
$\Phi=(A_\mu, F_{\mu\nu}, \phi)$ thus now
become
\[
\nabla^\alpha A_\alpha=0,
\]
\[
A_{(\mu,\nu)}=F_{\mu\nu},
\]
\[
\nabla^\alpha F_{\alpha\mu}=ie(\phi\overline{D_\mu\phi}-\overline\phi
D_\mu\phi),
\]
\[
g^{\mu\nu}D_\mu D_\nu\phi=0.
\]
Applying $L_K$ to these equations, and using
the equation $(\ref{kileq})$ yields
\[
\nabla^\alpha (L_KA)_\alpha={\bf L}(\nabla L_Kg)+{\bf L}(L_Kg),
\]
\[
(L_KA)_{(\mu,\nu)}={\bf L}(L_K\phi)+{\bf L}(L_Kg),
\]
\[
\Box (L_K\phi)={\bf L}(\nabla L_K(g))+{\bf L}(L_KA)+{\bf L}(L_Kg)
+{\bf L}(L_KF).
\]
Here, the notation ${\bf L}(x)$ means terms linear in $x$.
Noting that
\[
L_KT={\bf L}(g)+{\bf L}(\phi)+{\bf L}(A)+{\bf L}(F),
\]
and that $L_KF_{\mu\nu}=(L_KA)_{(\mu,\nu)}$,
we have that
given $g$, $A$, $F$, and $K$, the above system coupled with the equation
\[
\Box L_Kg={\bf L}(L_KT)+{\bf L}(L_Kg)
\]
can be written as a closed linear hyperbolic
system in $1+1$ dimensions for $L_KA$,
$(L_K\phi)$, and $(L_Kg)$, with
vanishing initial data on $H^+\cup H^-$, and for which
$0$ is a solution. Since $0$ is a solution 
of this system it must be the only solution, by uniqueness
of this initial value problem,
i.e.~the final assumption of Section $3$ is also verified.

The argument clearly also applies to the massive case,
and to more general so-called 
Higgs fields.

\end{document}